\documentclass{article}
\usepackage{amssymb,amsmath,amsthm,amsbsy,xcolor}
\usepackage{epsfig,amssymb,amsmath,textcomp,caption,verbatim,gensymb,wrapfig,mathtools, tabularx, algorithm2e,subfig,diagbox}
\usepackage{epstopdf, ltablex,booktabs,layouts}
\usepackage{graphicx}
\usepackage{amsfonts,braket}
\usepackage{geometry}
\usepackage[font={small}]{caption}
\usepackage{subfig}
\usepackage[colorlinks=true,citecolor=blue]{hyperref}
\usepackage{cleveref}
\usepackage[square,comma,numbers,compress]{natbib}

\begin{document}
	
\title{A consistent two-skyrmion configuration space from instantons}

\author{Chris Halcrow\footnote{email address: c.j.halcrow@leeds.ac.uk}\,\, and Thomas Winyard\footnote{email address:  t.winyard@leeds.ac.uk}\\ School of Mathematics, University of Leeds\\
	Leeds LS2 9JT, United Kingdom}

\maketitle

\abstract{To study a nuclear system in the Skyrme model one must first construct a space of low energy Skyrme configurations. However, there is no mathematical definition of this configuration space and there is not even consensus on its fundamental properties, such as its dimension. Here, we propose that the full instanton moduli space can be used to construct a consistent skyrmion configuration space, provided that the Skyrme model is coupled to a meson. Each instanton generates a unique skyrmion and we reinterpret the $8N$ instanton moduli as physical degrees of freedom in the Skyrme model. In this picture a single skyrmion has six zero modes and two non-zero modes: one controls the overall scale of the solution and one the energy of the meson field. We study the $N=1$ and $N=2$ systems in detail. Two interacting skyrmions can excite the meson field through scattering, suggesting that the meson and Skyrme fields are intrinsically linked. Our proposal is the first consistent manifold description of the two-skyrmion configuration space. The method can also be generalised to higher $N$ and thus provides a general framework to study any skyrmion configuration space. }

\section{Introduction}

The Skyrme model is a nonlinear theory of pions, where nuclei are described by skyrmions -- solitons whose existence and stability are due to the topology of the system \cite{Skyrme1961}. Each field configuration has a topological degree $N$ and this is identified with the baryon number of the corresponding nucleus. The energy minimising configuration with degree $N$ is called the $N$-skyrmion.

Skyrmions have been used with some success to describe light stable nuclei \cite{Manko:2007pr}, the nucleon-nucleon interaction \cite{VMLLCL1985, HH2020} and neutron stars \cite{Adam:2020yfv}. To describe nuclei, one usually starts with a space of classical Skyrme configurations and then applies a quantisation method. The simplest method only includes the zero modes of the skyrmion, physically allowing it to rigidly rotate (and isorotate). In reality, skyrmions can deform: they elongate \cite{Battye:2014qva}, change from one geometric shape into another \cite{Rawlinson:2017rcq} and break into clusters \cite{Gudnason:2018ysx}. These effects have been included in a few special situations \cite{Halcrow:2015rvz,Halcrow:2016spb}, but there is no systematic framework to model them. To develop such a framework, we must have a good understanding of the full $N$-skyrmion configuration space. This should include all low energy configurations but beyond this, little is known about the space.

In soliton systems, one can use a Bogomolny argument to bound the static energy of a degree $N$ soliton by a multiple of $N$. If this bound can be saturated the system is called BPS, and the second order Euler-Lagrange equations can be reduced to a set of first order BPS equations. For a given degree $N$, all solutions of the first order equations have the same energy, and form the $N$ moduli space. The resulting space is usually a differentiable manifold and we can ask the important question: what is the dimension of this manifold? For BPS systems, there is often a simple answer: if the dimension of the $1$-moduli space is $k$, then the dimension of the $N$-moduli space is $Nk$. This result can often be proved using the index theorem, introduced by Atiyah and Singer \cite{AS1963}.

Non-BPS systems, including the Skyrme model, are more complicated. We want to generate a configuration space of low energy Skyrme fields, but there is no mathematical criteria for which configurations to include and which to exclude. For BPS systems, we can use the BPS equations. For skyrmions, we must rely on physical intuition. A single skyrmion has six moduli: three translations and three orientations. The $2$-skyrmion configuration space should contain degrees of freedom which describe two well separated skyrmions, and hence it should be at least 12-dimensional. Recently, it was shown that two skyrmions with equal initial sizes can change their relative size after scattering \cite{Halcrow:2020lez}. Hence size fluctuations are low energy excitations and should be included in the configuration space. These additional degrees of freedom increase the dimension of the configuration space to 14. This agrees with numerical work by Gudnason--Halcrow, who used numerical techniques to find 14 low energy vibrational modes of the $N=2$ skyrmion \cite{Gudnason:2018ysx} and approximately $7N$ modes of skyrmions up to $N=8$.

However, it has long been known that skyrmions are well approximated by holonomies of instantons, first discovered by Atiyah and Manton \cite{Atiyah:1992if}. Sutcliffe, inspired by the Sakai-Sugimoto model \cite{SS2004}, showed that instantons generate approximate solutions to the Skyrme model coupled to an infinite tower of vector mesons \cite{Sutcliffe:2010et, Sutcliffe:2011ig}. The $N$-instanton moduli space is $8N$ dimensional, in disagreement with the $7N$ modes found numerically. What explains this discrepancy? Physically, what do the additional modes describe? In this paper we show that these questions are resolved by the addition of vector mesons to the Skyrme model. The additional degrees of freedom then control the energy of the meson fields, relative to the Skyrme field. 

Our result resolves several longstanding problems in the Skyrme model. First, we resolve the disagreement in dimension counting between Sutcliffe and Gudnason--Halcrow. We propose that solutions of the Skyrme model coupled to a single vector meson have $8N$ low energy modes. If the meson is removed, the leftover skyrmions will only have $7N$ modes. Most importantly, we show that the Skyrme model coupled to a single vector meson can be understood using the full instanton moduli space. When the Skyrme model has no vector mesons, the map between instantons and the $2$-skyrmion configuration space is not injective. This issue may appear in different guises. For example, the 12-dimensional $2$-skyrmion space proposed by Atiyah and Manton in \cite{Atiyah:1992if} is not a manifold, and is hence difficult to deal with using standard quantisation techniques. Our construction provides a consistent manifold description of the two-skyrmion configuration space: the first of its kind.

Our method can be applied more broadly since the $N$ skyrmion configuration space can be modelled by the $N$ instanton moduli space. Hence our proposal is the first step in a systematic framework allowing the study of any nucleus using instanton data.

We will demonstrate the central idea of this paper using the $N=1,2$ systems. Specifically we show that a consistent 16-dimensional $2$-skyrmion configuration space can be constructed. This contains the asymptotic region describing well separated skyrmions, which is required for quantum scattering \cite{Halcrow:2020lez}. To solve the quantum two skyrmion problem one then needs to construct this space in detail and solve Schr\"odinger's equation on it. While computationally complex, no major theoretical roadblocks remain: there is now a roadmap for how to study the full nucleon-nucleon problem in the Skyrme model.

The paper is structured as follows. We first introduce the instanton-generated Skyrme model, coupled to an infinite tower of vector mesons, in section 2. We then truncate the infinite tower to include only one meson. Using the $N$ instanton moduli space, we generate Skyrme and meson fields and interpret their physical degrees of freedom. We do this for the $N=1$ sector in section 3 and the $N=2$ sector in section 4. Concluding remarks and future work is contained in section 5.

\section{The model}

We study the Skyrme model coupled to a tower of vector mesons, with the interactions fixed by a holographic construction. We now review this construction, closely following \cite{Sutcliffe:2011ig}. We begin with $SU(2)$ Yang-Mills theory in $\mathbb{R}^4$. We often single out the holographic direction, and as such write $\mathbb{R}^4 = \mathbb{R}^3_{\boldsymbol{x}}\times\mathbb{R}_h$ to help clarify the calculation. The Yang-Mills energy functional is
\begin{equation} \label{eq:YM}
E_{YM} = -\frac{1}{8}\int \text{Tr}\left( F_{IJ}F_{IJ} \right)d^3\boldsymbol{x} \, dh \, ,
\end{equation}
where $I = 1,2,3,4$ and $(\boldsymbol{x}, h)$ denote the spatial and holographic directions in $\mathbb{R}^4$, $F_{IJ} = \partial_IA_J - \partial_JA_I + [A_I,A_J]$ is the field strength and $A\in \mathfrak{su}(2)$ is the gauge field. We use upper case latin indices to denote coordinates on $\mathbb{R}^4$ and lower case latin indices for spatial coordinates on $\mathbb{R}^3_{\boldsymbol{x}}$, so that $A_I = (A_i, A_h)$.  This model has solutions called instantons classified by the degree of the map $A$, usually called the instanton number $N\in\mathbb{Z}$. For each $N$, there is a moduli space of $8N$ instantons, which all have energy $E_{YM} = 2\pi^2 N$. 

To connect the four-dimensional Yang-Mills theory to the three-dimensional Skyrme model, we deconstruct the holographic dimension of the theory. To do so, we must switch to the temporal gauge $A_h = 0$. Provided that $A_I \to 0$ as $|h| \to \infty$, we can perform this gauge transformation $A_I \to gA_I g^{-1} - \partial_I g g^{-1} $ using $g \in SU(2)$ defined by
\begin{align} \label{eq:gaugeeq}
gA_h- \partial_h g = 0 \, .
 \quad 
\end{align}
In this gauge, $A$ is related to the Skyrme field $U \in SU(2)$\footnote{Specifically through the right current $R_i = \partial_i U U^{-1}$.} and the meson fields $V^n \in \mathfrak{su}(2)$ by a mode decomposition
\begin{equation} \label{eq:Ainpsi}
	A_i = -\partial_i U U^{-1} \psi_+(h) + \sum_{n=0}^\infty V_i^n(\boldsymbol{x})\psi_n(h) \, .
\end{equation}
There is some freedom in the choice of basis functions $\psi$. Sutcliffe chose a basis to make the theory with a single meson as close to BPS as possible,
\begin{equation} \label{eq:basis}
	\psi_+ = \frac{1}{2}(1+\text{erf}(h/\sqrt{2})) \, , \quad \psi_n = \frac{(-1)^n}{\sqrt{n!2^n \sqrt{\pi}}}e^{h^2/2}\frac{d^n}{dh^n}e^{-h^2} \, .
\end{equation}
Different choices of basis functions will lead to different interactions between the skyrme and meson fields, and one could try to use nuclear data to help fix these functions. Our work is qualitative and it is helpful to compare our results to those of \cite{Sutcliffe:2011ig, Naya:2018kyi,Naya:2018mpt} so we simply adopt their choice \eqref{eq:basis}.  We write the Skyrme and meson fields in terms of the gauge field by inverting equation \eqref{eq:Ainpsi}:
\begin{equation} \label{eq:makesk}
\partial_i U U^{-1} = -A_i(\infty,\boldsymbol{x}), \quad 
V_i^n = \int_{-\infty}^\infty\left( A_i(t,\boldsymbol{x}) + R_i(\boldsymbol{x})\psi_+(t) \right)\psi_n(t) \, \text{d}t \, .
\end{equation}

Substituting \eqref{eq:Ainpsi} into the Yang-Mills energy \eqref{eq:YM} and doing the $h$-integral gives the energy functional of our theory. We will now restrict the theory to only include one vector meson. Since there is a unique meson field we write this as $V \equiv V^0$. The energy can then be written in three parts: the Skyrme energy, the meson energy and the interaction energy. We denote these as
\begin{equation} \label{eq:toteng}
	E = E_S + E_V + E_I  = \int \left( \mathcal{E}_S + \mathcal{E}_V + \mathcal{E}_I  \right)\, d^3\boldsymbol{x} \, , 
\end{equation}
where
\begin{align} \label{eq:energies}
	\mathcal{E}_S &=
	\text{Tr}\left( \frac{c_1}{2}R_i R_i+\frac{c_2}{16}[R_i,R_j][R^i,R^j] \right) \\
	\mathcal{E}_{\rm V}&= \text{Tr}\left(
	\frac{1}{8}(\partial_i V_j-\partial_j V_i)^2
	+\frac{1}{4}m^2V_i^2
	+c_3(\partial_i V_j-\partial_j V_i)[V^i,V^j]
	+c_4[V_i,V_j]^2
	\right)  \\
	\mathcal{E}_I &= \text{Tr}\bigg(
	c_5[R_i,R_j](\partial_i V_j - \partial_j V_i)
	-c_6[R_i,R_j][V_i,V_j]
	-c_7[R_i,R_j][V_i,V_j]\nonumber\\
	&
	+\frac{1}{2}c_6[R_i,R_j]([R_i,V_j]-[R_j,V_i])
	-\frac{1}{8}([R_i,V_j]-[R_j,V_i])(\partial_i V_j-\partial_j V_i)
	\nonumber\\
	&-\frac{1}{2}c_{3}([R_i,V_j]-[R_j,V_i])[V_i,V_j]
	\bigg),
\end{align}
and the constants are given by\footnote{These have analytic expressions in terms of integrals of the basis functions $\psi_n$. See \cite{Sutcliffe:2011ig} for details.}
\begin{align}
c_1 = 0.141, \quad c_2 = 0.198, \quad c_3 = 0.153, \quad c_4 = 0.050, \\
c_5 = 0.038, \quad c_6 = 0.078, \quad c_7 = 0.049, \quad m = 0.707 \, .
\end{align}
 The instanton-generated fields given by \eqref{eq:makesk} provide a good approximation to the low energy configurations of \eqref{eq:toteng}. In the rest of the paper, we aim to give a physical understanding of the instanton-generated fields in terms of instanton solutions.

\section{The N=1 instanton-generated skyrmion}

The $N=1$ instanton has spherical symmetry in $\mathbb{R}^4$ which we can exploit to study the solution in detail. To begin, consider the centred BPST instanton given by \cite{Belavin:1975fg}
\begin{equation}
	A_\text{BPST} =  \left( -i\frac{ x_b\tau_b}{ h^2 + r^2 + \lambda^2},   -\frac{\epsilon_{bij}x_j }{ h^2 + r^2 + \lambda^2}\tau_b + i\delta_{ib}\frac{h}{ h^2 + r^2 + \lambda^2 }\tau_b \right) \, ,
\end{equation} 
where $r = |\boldsymbol{x}|$ and $\tau_a$ are the Pauli matrices. To use the expressions from section 2, we must first apply a gauge transformation $A \to \tilde{A}$ to enter the temporal gauge $\tilde{A}_h = 0$. The solution of \eqref{eq:gaugeeq} is given by
\begin{align} \label{eq:Fdef}
	g = &\mathcal{P} \exp\left( \int_{-\infty}^h A_h\left( \boldsymbol{x}, h' \right) dh' \right) \nonumber \\
	= &\exp \left( - \frac{i x_i\tau_i}{2\sqrt{r^2 + \lambda^2} }\left( \pi + 2\tan^{-1}\left(\frac{h}{\sqrt{r^2+\lambda^2}} \right) \right) \right) \nonumber \\ \equiv &\exp i F(r,h) \tau_i \hat{x}_i \, .
\end{align}
Applying this gauge transformation to $A_\text{BPST}$ we find the centred $N=1$ instanton in the temporal gauge
\begin{align} \label{1inst}
	\tilde{A}_i( \boldsymbol{x},h) = \,&g A_i g^{-1} - (\partial_i g) g^{-1} \nonumber \\
		= &\Bigg((\delta_{ia} - \hat{x}_i\hat{x}_a)\left( \eta \cos 2F - \left(\zeta + \frac{1}{2r} \right) \sin 2F \right) + \hat{x}_i\hat{x}_a \left( \eta -  \partial_r F \right)\nonumber\\ &+ \epsilon_{ija}\hat{x}_j\left( \eta \sin 2F + \left(\zeta + \frac{1}{2r} \right) \cos 2F - \frac{1}{2r} \right)  \Bigg) \tau_a \, , 
\end{align} 
where
\begin{align}
	\eta = \frac{h}{ h^2 + r^2 + \lambda^2} \quad \text{and} \quad \zeta = -\frac{r}{ h^2 + r^2 + \lambda^2} \, .
\end{align}
This solution contains a single moduli: the size $\lambda\in\mathbb{R}_{>0}$. The remaining seven moduli of the instanton are generated by acting on $\tilde{A}$ with translational and global gauge symmetries. Doing so, we find the family of solutions
\begin{equation} 
	A(\boldsymbol{X},H, G;  \boldsymbol{x},h) = G\tilde{A}(\boldsymbol{x} - \boldsymbol{X},h-H)G^{-1} \, ,
\end{equation}
where $(\boldsymbol{X},H)\in\mathbb{R}^3_{\boldsymbol{x}}\times \mathbb{R}_h$ give the position of the instanton and $G \in SU(2)$ the global gauge orientation. Overall the eight-dimensional moduli space is isomorphic to $\mathbb{R}_{>0} \times SU(2) \times \mathbb{R}^4$. Substituting \eqref{1inst} into \eqref{eq:toteng} we find that the instanton-generated Skyrme field is given by
\begin{equation} \label{eq:skyprof}
	U = g(\boldsymbol{x}, \infty) = \exp \, i F(r,\infty)\tau_i \hat{x}_i =  \exp\left( - i \frac{ r }{  \sqrt{\lambda^2 + r^2} } \hat{x}_i \tau_i \right) \, ,
\end{equation}
where $F(r,\infty)$ is the profile function of the $N=1$ skyrmion with spherical symmetry. The vector mesons are given by
\begin{align} \label{eq:mesprof}
	V_i(\boldsymbol{x}) &=\left( (\delta_{ia} - \hat{x}_i\hat{x}_a)k_1(r) + \hat{x}_i\hat{x}_ak_2(r) + \epsilon_{iaj}\hat{x}_jk_3(r)\right)\tau_a \, ,
\end{align}
where
\begin{align} \label{eq:mesdef}
	k_1 &= \int_{-\infty}^\infty \left( \eta \cos 2F - \left(\zeta + \frac{1}{2r} \right) \sin 2F  +(  \sin 2f/ 2r )\psi_+ \right)\psi_0\, dh \\
	k_2 &= \int_{-\infty}^\infty \left(   \eta -  \partial_r F + (  \partial_r f )\psi_+   \right) \psi_0 \,dh \\
	k_3 &= \int_{-\infty}^\infty \left( \eta \sin 2F + \left(\zeta + \frac{1}{2r} \right) \cos 2F - \frac{1}{2r} 
	-  \left( \frac{\cos 2f }{ 2r} - \frac{1}{2r} \right)\psi_+ \right) \psi_0 \, dh \, .
\end{align}
Note that the $k_i$ depend on $H$ through $\eta(h-H)$ and $\zeta(h-H)$. This is a small generalisation of the results in \cite{Sutcliffe:2010et}. There, the gauge field had definite parity whereas ours does not.

We can use the fields \eqref{eq:skyprof} and \eqref{eq:mesprof} to study the energy of the instanton-generated Skyrme and meson configurations as a function of the moduli space parameters. There are  two parameters which affect the energy: $\lambda$ and $H$. We list the energy \eqref{eq:toteng} for a variety of $\lambda$ and $H$ in Table \ref{tab:Ealam}. As the energy is even in $H$ we take $H\geq0$. The minimum of the energy is at $\lambda = 1.2$ and $H=0$, confirming the result of Sutcliffe \cite{Sutcliffe:2010et}.

\begin{table}
	\centering
	\begin{tabular}{ c | c  c c c c c }  
		\backslashbox{$\lambda$}{$H$} & 0.0 & 0.2 & 0.4 & 0.6 & 0.8 & 1.0 \\ \hline
		0.8 & 1.20176 & 1.20167 & 1.21298 & 1.26394 & 1.38399 & 1.5884 \\
		1.0& 1.09693 & 1.10212 & 1.12397 & 1.17798 & 1.28065 & 1.44099  \\
		1.2 & 1.07114 & 1.07863 & 1.10459 & 1.15772 & 1.24744 & 1.37901 \\
		1.4 & 1.08809 & 1.09645 & 1.12349 & 1.17417 & 1.25383 & 1.36544  \\
		1.6 &1.13023 & 1.13876 & 1.16545 & 1.2131 & 1.2847 & 1.3818  \\
	\end{tabular}
\caption{Energies of \eqref{eq:toteng} for a variety of $H$ and $\lambda$. Energies are in units of $2\pi^2$. In these units, a solution which saturates the BPS bound has $E = 1$. }
	\label{tab:Ealam}
\end{table}

We would like to exmaine how the skyrme and meson fields change as $\lambda$ and $H$ are varied. However, it is not obvious how to seperate the skyrme and meson contributions as the fields are strongly coupled. This is related to the issue of gauge invariance. Applying a gauge transformation to the instanton gauge field $A$ modifies the skyrme-meson Lagrangian \eqref{eq:toteng} and the solutions \eqref{eq:mesprof}. This can be seen explicitly in \cite{Sutcliffe:2010et} and \cite{Sutcliffe:2011ig}, where the $N=1$ calculation is done in the definite-parity and a no-parity gauge. The calculation there suggests that the Skyrme energy $\mathcal{E}_S$ and the remaining energy $\mathcal{E}_V + \mathcal{E}_I$ are both gauge invariant, while $\mathcal{E}_V$ is not. Hence we will use $\mathcal{E}_S$ and $\mathcal{E}_V + \mathcal{E}_I$ as proxies for how each field changes.

\begin{figure}[h!]
	\centering
		\includegraphics[width=0.8\textwidth]{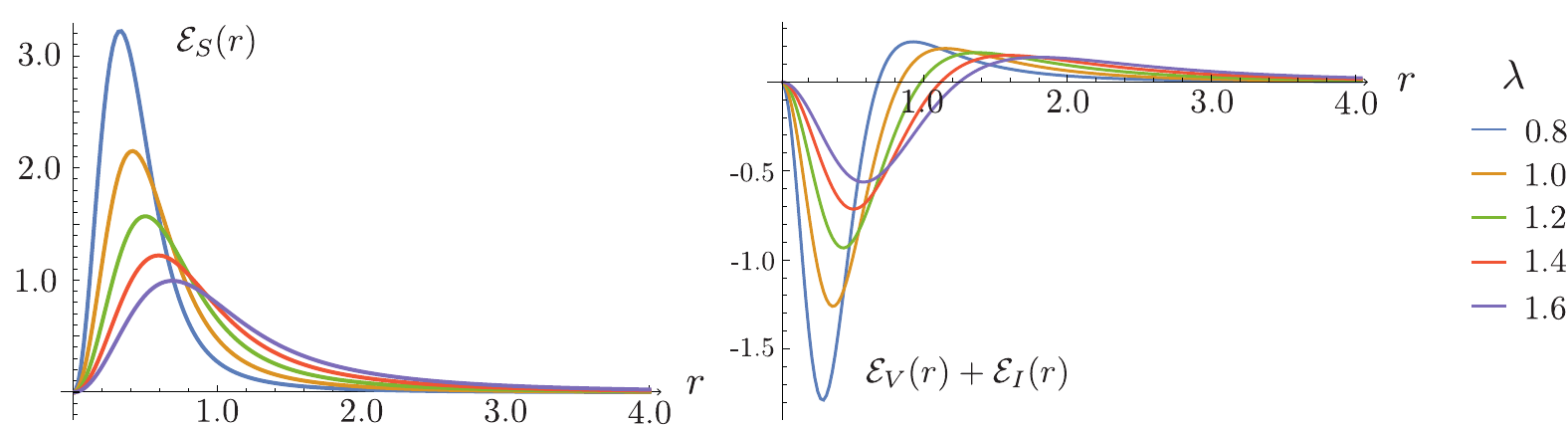} 
	\caption{A plot of the Skyrme energy density $\mathcal{E}_S(r)$ and remaining energy density $\mathcal{E}_V(r)+\mathcal{E}_I(r)$  for $\lambda = 0.8 \to 1.6$.}
	\label{fig:FVlam}%
\end{figure}

It is known that $\lambda$ controls the size of the Skyrme field. This can be seen in a variety of ways. For instance, the profile function satisfies $F_\lambda(r, \infty) \approx F_{\tilde{\lambda}}(\mu r, \infty)$ for some  $\bar{\lambda}, \mu\in\mathbb{R}$. We can also examine the radial energy density $\mathcal{E}_S(r)$, defined as $E_S = \int \mathcal{E}_S(r) \, dr$. We plot this quantity for a variety of $\lambda$ in Figure \ref{fig:FVlam}, and see that as $\lambda$ grows the energy density becomes concentrated at larger values of $r$. We also plot $\mathcal{E}_V(r) + \mathcal{E}_I(r)$ for a variety of $\lambda$ and see that a similar phenomena occurs. From this we deduce that $\lambda$ controls the size of both the skyrme and meson fields.

\begin{figure}[h!]
	\centering
	\includegraphics[width=0.5\textwidth]{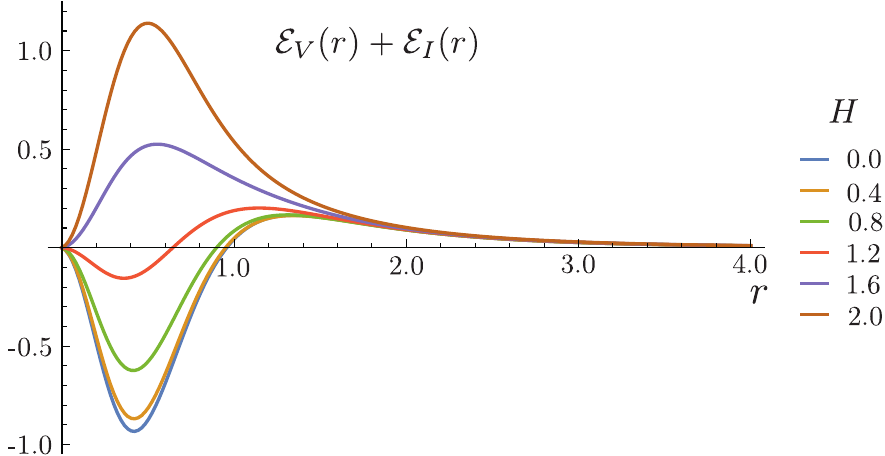} 
	\caption{The energy density $\mathcal{E}_V(r)+\mathcal{E}_I(r)$ for $\lambda = 1.2$ and $H \in [0,2]$.}
		\label{fig:FVa}
\end{figure}

Now consider varying $H$. Since the Skyrme profile function $F$ is independent of $H$, the Skyrme energy $\mathcal{E}_S(r)$ does not change. In contrast the meson profiles $k_i$ do depend on $H$ and so the energy density $\mathcal{E}_V(r) + \mathcal{E}_I(r)$ does change. We plot this density for a variety of $H$ and a fixed $\lambda = 1.2$ in Figure \ref{fig:FVa}. The energy density is even in $H$ and so we fix $H\geq0$. As $H$ increases, the meson energy density grows at every point. Hence this mode simply injects energy into the meson field. The minimal energy configuration occurs at $\lambda=1.2$ and $H=0$.

In summary, a skyrmion coupled to a meson field has eight degrees of freedom represented by the instanton moduli. There are six zero modes: translations (controlled by $\boldsymbol{X}$) and rotations ($G$). There are two non-zero modes: the overall size ($\lambda$) and the energy of the meson field ($H$).

\subsection{Mode analysis}

The newly discovered mode, modelled by changing $H$, is a zero mode of the instanton. As such, it should correspond to a zero mode of the Skyrme model coupled to the infinite tower of vector mesons. Let us briefly reintroduce the infinite tower of mesons. The zero mode (at $H=0$) is given by
\begin{align}
	\frac{\partial V_i^n(\boldsymbol{x}) }{\partial H}\bigg\rvert_{H=0} &= -\int_{-\infty}^{\infty} \partial_H(A_i(h-H))\rvert_{H=0} \psi_n(h) \, dh \\
	&= \int_{-\infty}^{\infty} A_i(\boldsymbol{x}, t)\rvert_{H=0} \partial_h\psi_n(h) \, dh \, ,
\end{align}
where we have swapped the $H$ derivative for an $h$ derivative, then integrated by parts. We can calculate these in the spherically symmetric case, finding
\begin{align}
	\frac{\partial V_i^n }{\partial H}\bigg\rvert_{H=0} &= \int_{-\infty}^{\infty} A_i(\boldsymbol{x}, t) \partial_h\psi_n(h) \, dh \\
	&= (\delta_{ia} - \hat{x}_i\hat{x}_a)\kappa^n_1(r) + \hat{x}_i\hat{x}_a\kappa^n_2(r) + \epsilon_{iaj}\hat{x}_j \kappa^n_3(r) \, ,
\end{align}
where
\begin{align}
	\kappa^n_1(r) &= \int_{-\infty}^{\infty}\left (\eta \cos(2F) - \left( \zeta +\frac{1}{2r} \right) \sin 2F \right)\Big\rvert_{H=0} \partial_h\psi_n(h) \, dh\\
	\kappa^n_2(r) &= \int_{-\infty}^{\infty}\left (  \eta - \partial_r F   \right)\Big\rvert_{H=0} \partial_h\psi_n(h) \, dh \\
	\kappa^n_3(r) &= \int_{-\infty}^{\infty}\left( \eta \sin 2F + \left(\zeta + \frac{1}{2r}\right) \cos 2F - \frac{1}{2r}    \right)\Big\rvert_{H=0} \partial_h\psi_n(h) \, dh \, .
\end{align}
This zero mode allows the infinite tower of mesons to exchange matter with no cost. The transformation is likely related to the enhanced local hidden symmetry present in the holographic model. When there is only one vector meson the mode does cost significant energy, as can be seen in Table \ref{tab:Ealam}.

\subsection{Numerical method}

Generally, there is no spherical symmetry when $N>1$ and we cannot generate analytic expressions for the fields. Instead we must generate the skyrmions numerically. Given an instanton configuration we first discretise $\mathbb{R}^3_{\boldsymbol{x}}$ evenly. We must then apply a gauge transform $A_I \to \tilde{A}_I =  gA_I g^{-1} - \partial_I g g^{-1} $ which takes $A$ into the temporal gauge. Hence, we seek a local gauge transformation $g(\boldsymbol{x},h)$ at each spatial point which satisfies  \eqref{eq:gaugeeq}. We follow \cite{Leese:1993mc} and write
\begin{equation}
	g = g_0 + i g_i \tau_i \quad A_h =i \alpha_i \tau_i \, ,
\end{equation}
which reduces \eqref{eq:gaugeeq} to 
\begin{align} \label{eq:loceq}
	\partial_h g_0 = - \alpha_i g_i \quad \partial_h g_i = g_0 \alpha_i + \epsilon_{ijk}\alpha_jg_k\, , \quad g(\boldsymbol{x}, -\infty) = 1 \, .
\end{align}
To find $g$ we must numerically solve the initial value problem in \eqref{eq:loceq} on the interval $h \in [-\infty, \infty ]$. To simplify this, we introduce the coordinate $\tilde{h} = \arctan(h)$ and discretise the interval $[-\pi/2,\pi/2]$ evenly. This generates $\tilde{A}$ at each point which  is then used to calculate the Skyrme and meson fields using \eqref{eq:makesk}.

\begin{figure}[h!]
	\begin{center}
		\includegraphics[width=0.2\textwidth]{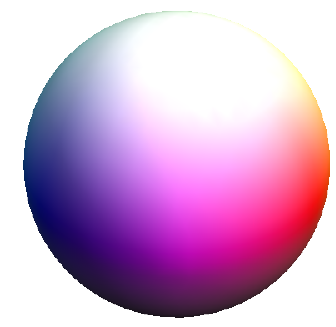} 
	\end{center}
	\caption{An energy density isosurface of a $B=1$ skyrmion, coloured as described at the start of section 4.1.} 
	\label{fig:b1}%
\end{figure}

Once we have the Skyrme field $U$ and meson fields $V$ we can then calculate the energy densities, baryon density and plot the configuration. We plot an energy isosurface and colour it depending on the value of the Skyrme field at each point on the surface. If we write $U = u_0 + i u_a \tau_a$ then we colour the density using the Runge colour sphere applied to $u_a$. This means the surface is white/black when $u_3 = \pm 1$, and red, blue and green when $u_1 + i u_2$ has argument $0$, $2\pi/3$ and $4\pi/3$ respectively. The $N=1$ instanton-generated skyrmion is plotted in figure \ref{fig:b1}. To ensure our numerical code is accurate, we generated this solution using BPST, t'Hooft and JNR data. As JNR data decays slowly, the skyrmion generated from an $N$-instanton has charge $-N$. To remedy this, one can simply multiply the final Skyrme field $U$ by $-1$ \cite{Leese:1993mc}.

To ensure that we can trust our numerical techniques, we reproduced the $N=1$ results from table \ref{tab:Ealam} with an accuracy of two decimal places. To achieve this, we used a three-dimensional spatial grid of volume $10^3$ with $90^3$ lattice points and 500 time points for the holographic coordinate $\tilde{h}$. We also extracted the Skyrme and vector meson profile functions and compared them to the analytical expressions \eqref{eq:Fdef} and \eqref{eq:mesdef}.

\section{The N=2 instanton-generated skyrmions}

\begin{figure}[h!]
	\begin{center}
		\includegraphics[width=0.2\textwidth]{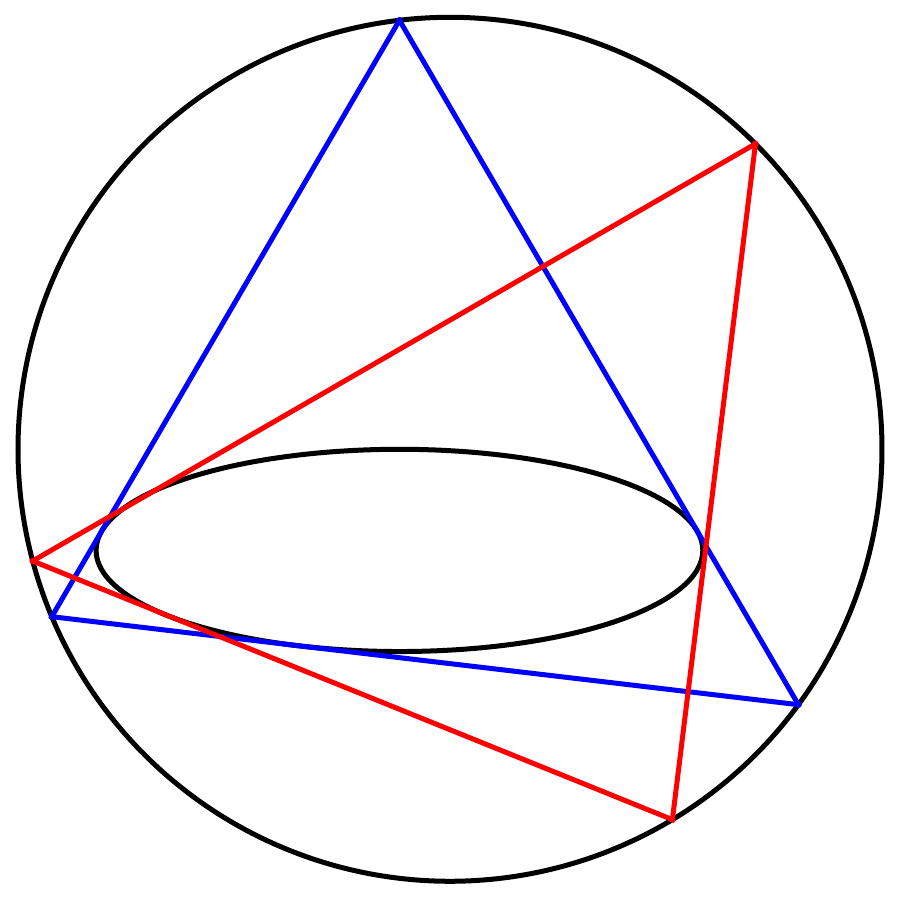} 
	\end{center}
	\caption{Two triangles related by the Poncelet condition.} 
	\label{fig:poncelet}%
\end{figure}

All $N=2$ instantons are described by the JNR ansatz \cite{Jackiw:1976fs}. The gauge field is given by
\begin{equation}
	A_\mu = \frac{i}{2} \sigma_{\mu \nu} \partial_\nu \log(\rho) \, ,
\end{equation}
where $\sigma_{\mu \nu}$ is antisymmetric and defined by
\begin{equation}
	\sigma_{i4} = \tau_i \quad \sigma_{ij} = \epsilon_{ijk}\tau_k
\end{equation}
and the density $\rho$ is given by
\begin{equation}
	\rho = \frac{\lambda_1}{|x-X_1|^2} + \frac{\lambda_2}{|x-X_2|^2}+ \frac{\lambda_3}{|x-X_3|^2} \, ,
\end{equation}
where $X_i\in\mathbb{R}^4$ and $\lambda_i \in \mathbb{R}_{>0}$ define the JNR data. The vectors $X_1, X_2, X_3$ form a triangle. Hartshorne simplified the geometry of the data using Poncelet's porism \cite{Hartshorne:1978vv}. Consider a triangle inscribed by an ellipse and circumscribed by a circle. Poncelet showed that this is just one of a one-dimensional family of triangles, which are similarly related to the ellipse and circle - an example is shown in figure \ref{fig:poncelet}. Hartshorne showed that JNR data related by Poncelet's condition gives rise to instantons which are equivalent up to a global gauge transform. Hence, to describe instantons up to gauge equivalence, one can specify the JNR data by specifying a circle and an ellipse. The common symmetries of the circle and ellipse are inherited by the corresponding instanton. The $\lambda_i$ are fixed by considering the points on the triangle tangent to the ellipse. Denote the point opposite $X_i$ as $Y_i$, then
\begin{equation}
	\frac{\lambda_1}{\lambda_2} = \frac{|X_1-Y_3|}{|X_2-Y_3|}, \quad \frac{\lambda_2}{\lambda_3} = \frac{|X_2-Y_1|}{|X_3-Y_1|}, \quad \frac{\lambda_3}{\lambda_1} = \frac{|X_3-Y_2|}{|X_1-Y_2|} \, .
\end{equation} 
Only ratios of the $\lambda_i$ are physically relevant, so we can fix $\lambda_1$ to any non-zero constant, such as $\lambda_1 = 1$.

\begin{figure}[h!]
	\begin{center}
		\includegraphics[width=0.8\textwidth]{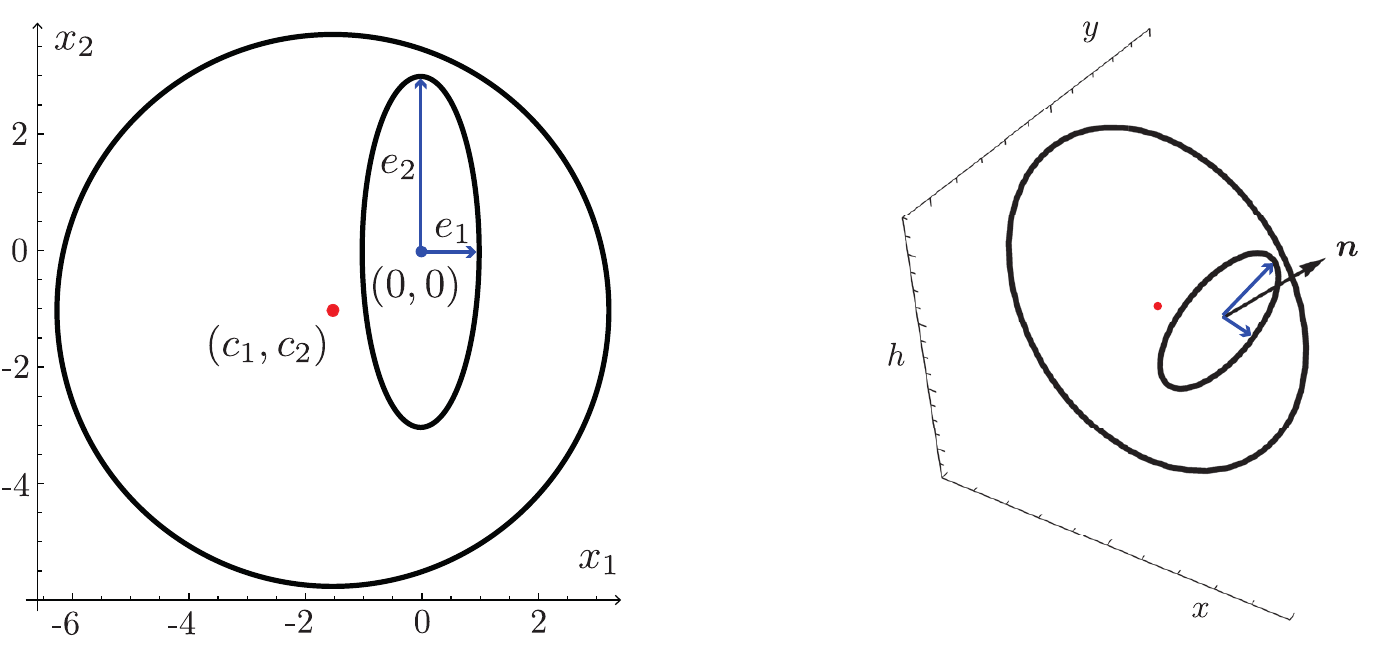} 
	\end{center}
	\caption{A configuration with $(e_1,e_2) = (1,3)$, $(c_1, c_2) = (-3/2,-1)$ and $\boldsymbol{n} = (1,0,1 )/\sqrt{2}$. The circle and ellipse lie in the $2$-plane with normal $\boldsymbol{n}$. The coordinate $x_1,x_2$ form a basis for the embedded two-plane.} 
	\label{fig:paras}%
\end{figure}

We can now describe any $N=2$ instanton by specifying a circle and an ellipse embedded on a 2-plane in $\mathbb{R}^4$. To describe how these parameters translate to the Skyrme model, we begin with a well understood configuration and consider deformations around it.  The centre of the ellipse takes a value in $\mathbb{R}^4$. The position and embedding of the 2-plane in $\mathbb{R}^3_{\boldsymbol{x}}$ determines the centre of mass and global rotational orientation of the skyrmions. Wlog we place the ellipse at the origin of $\mathbb{R}^4$ and place the 2-plane in the $z=0$ hyperplane. We then fix the JNR triangle by choosing $X_3$ parallel with the minor axis of the ellipse, like the blue triangle in figure \ref{fig:poncelet}. The remaining coordinates are the axes of the ellipse $(e_1,e_2)$, the centre of the circle $(c_1,c_2)$, and the normal of the plane $\boldsymbol{n} \in \mathbb{R}_x\times \mathbb{R}_y\times \mathbb{R}_h$. The radius of the circle is fixed by the Poncelet condition. An example of a configuration labelled using our coordinates is displayed in figure \ref{fig:paras}

\begin{figure}[h!]
	\begin{center}
		\includegraphics[width=0.5\textwidth]{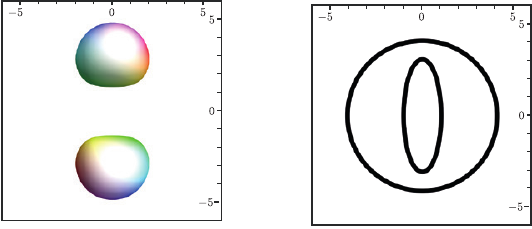} 
	\end{center}
	\caption{The energy density and JNR data for $(e_1,e_2,c_1,c_2) = (1,3,0,0)$. The $x$-axis is horizontal, while the $y$-axis is vertical. The centre of the energy density for each skyrmion is approximately positioned on the vertex of the ellipse.  } 
	\label{fig:initial}%
\end{figure}

We begin by setting $e_1=1, e_2=3$ and $c_1=c_2=0$ and placing the 2-plane in the $xy$-plane. The JNR data and the corresponding skyrmion are plotted in figure \ref{fig:initial}. The centre of each skyrmion is located at approximately the vertices of the ellipse and are orientated in the attractive channel (since their colours of closest contact match). This is the orientation where the skyrmions are maximally attractive. The JNR data from figure \ref{fig:initial} has a reflection symmetry in each coordinate of $\mathbb{R}^4$. Consider the symmetry group $G$ of the instanton generated skyrmion, which is the subgroup of the instanton symmetry group restricted to $\mathbb{R}^3_{\boldsymbol{x}}$. The symmetry group is generated by three reflection symmetries $x_i \to -x_i$. Skyrmion deformations can then be classified by the irreducible representations (irreps) of $G$. Since each generator has order two, we can classify these as $(n_1,n_2,n_3)$ where $n_i = \{+,-\}$.

\begin{figure}[h!]
	\begin{center}
		\includegraphics[width=1.0\textwidth]{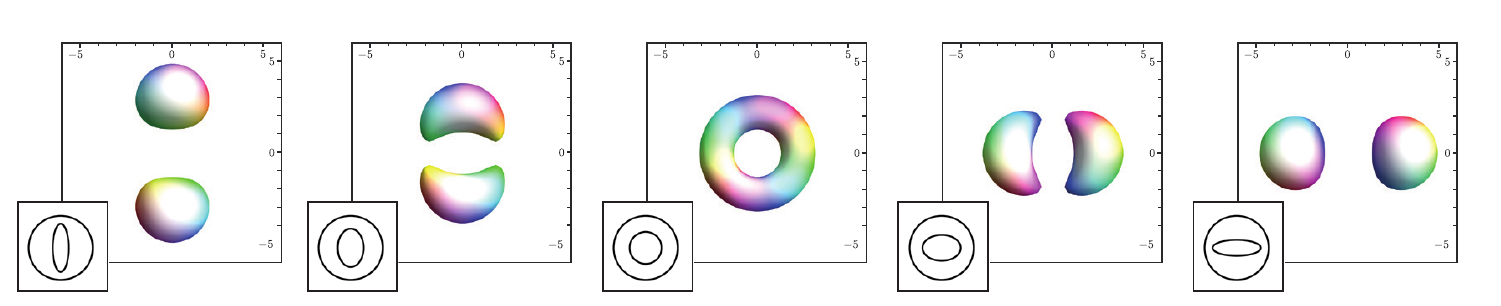} 
	\end{center}
	\caption{Attractive channel scattering for the family $F1$ \eqref{eq:t1}. We plot the energy density of the configuration, alongside the corresponding JNR data. The path is described in the text, and these plots correspond to $t_1= \{0,3\pi/8, \pi/2, 5\pi/8, \pi \}$. } 
	\label{fig:att}%
\end{figure}

There are two perturbations transforming as $(+++)$ which correspond to deformations of the ellipse. The first is an overall scalaing which controls the size of the skyrmion. The second modifies the shape of the ellipse. Consider the one-parameter family
 \begin{equation} \label{eq:t1}
F1: e_1 = 2-\cos(t_1), \quad e_2  =2+\cos(t_1) \, .
 \end{equation}
The skyrmions from the family of configurations $F1$ are plotted for $t_1\in[0,\pi]$ in figure \ref{fig:att}. We see that the skyrmions' positions are approximated by the vertices of the major axis of the ellipse. When $e_1=e_2$ the ellipse becomes a circle and there is no unique major axis. As such, the energy density is concentrated on a ring. This toroidal skyrmion is the global energy minimiser of the $N=2$ sector. The meson and Skyrme fields are concentrated in the same region of space for each $t_1$.  We will often discuss scattering skyrmions. By this we mean generating a path where the skyrmions pass through one another through a configuration where the ellipse becomes a circle, as is seen in figure \ref{fig:att}. We generate scatterings by continuously changing the major axis of the ellipse from lying on the $y$-axis to the $x$-axis. 

\begin{figure}[h!]
	\begin{center}
		\includegraphics[width=1.0\textwidth]{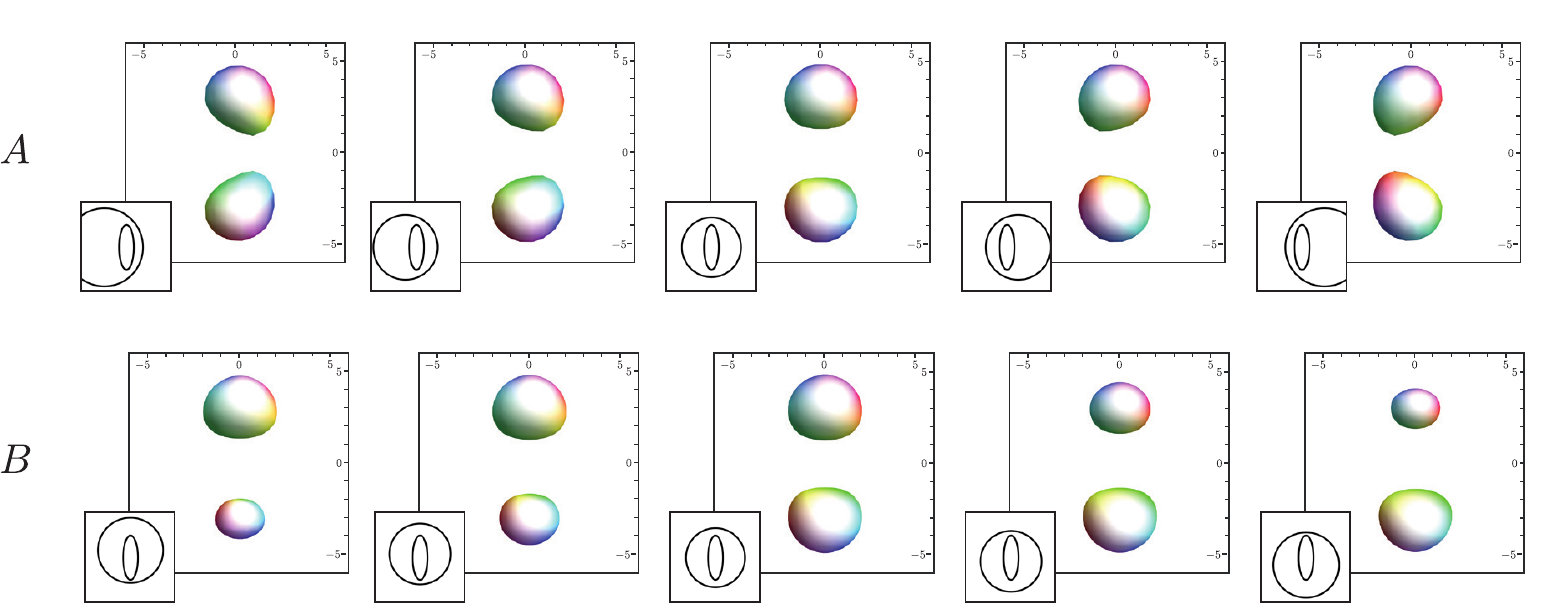} 
	\end{center}
	\caption{Moving the centre of the circle in the JNR data. In figure 8A, we plot configurations $F2$ \eqref{7a} with $t_2 = \{-3, -1.5, 0, 1.5, 3\}$. In figure 8B, we plot configurations $F3$ \eqref{7b} with $t_3 = \{-1,-3/4,0,3/4,1\}$.  } 
	\label{fig:circ1}%
\end{figure}

Next, consider changing the centre of the circle. If we fix the circle center wrt the minor axis of the ellipse ($c_2=0$) but modify wrt to the major axis of the ellipse ($c_1=0$), the Skyrme field retains a reflection symmetry in the $x$-axis.  This perturbation transforms as $(+-+)$. Physically, the skyrmion moves out of the attractive channel, as seen in figure \ref{fig:circ1}A. This is a zero mode when the skyrmions are infinitely separated but costs energy when they are close. Now allow $c_2$ to change but fix $c_1=0$. This perturbation transforms as $(-++)$ and corresponds to a relative size fluctuation of the skyrmions.  This can be seen in figure \ref{fig:circ1}B. In detail these two families are given by
\begin{align} 
	&F2: e_1=1, \quad e_2=3, \quad c_1 = -1 + t_2,\quad  c_2 = 0 \label{7a} \\
	&F3: e_1=1, \quad e_2=3, \quad c_1 = 0,  \quad c_2 = -1 + t_3 \label{7b} \, .
\end{align}

\begin{figure}[h!]
	\begin{center}
		\includegraphics[width=1.0\textwidth]{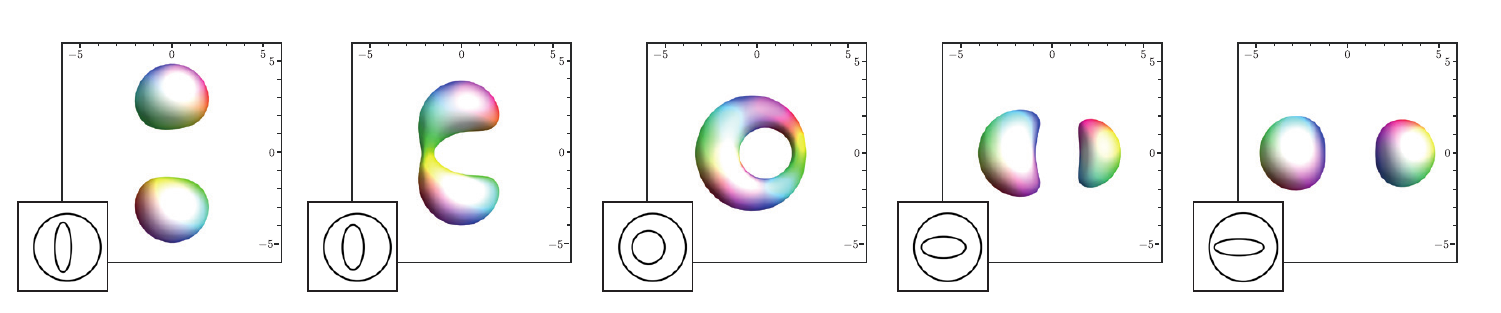} 
	\end{center}
	\caption{A scattering with ingoing skyrmions out of phase. The skyrmions perform 90$^\circ$ scattering, briefly merging into a deformed torus. The outgoing skyrmions have different sizes. The configurations are taken from $F4$ \eqref{eq:sizesca} with $t_4 = \{ 0,3\pi/8,\pi/2,5\pi/8,\pi \}$.  } 
	\label{fig:circ2}%
\end{figure}

The $(+-+)$ and $(-++)$ perturbations are related. If we begin by moving the skyrmions out of the attractive channel (by applying the $(+-+)$ perturbation) and then scatter them, the outgoing skyrmions have different sizes (the physical consequence of the $(-++)$ perturbation). This type of scattering is demonstrated by the family
\begin{equation} \label{eq:sizesca}
	F4: e_1 = 2 - \cos(t_4), \quad e_2 = 2 + \cos(t_4), \quad c_1 = 1/2, \quad c_2 = 0 \, .
\end{equation}
As discussed with $F1$, we have chosen a path where the ellipse at $t_4=0$ is a circle. We plot the configurations from $F4$ in figure \ref{fig:circ2}. Physically, a relative orientation can become a relative size fluctuation and vice versa. This suggests that size fluctuations are low energy excitations and should be included in any two-skyrmion configuration space. An analogous mode can be seen in the baby Skyrme model, where it can be understood in terms of rational map solutions of the nonlinear sigma model \cite{Halcrow:2020lez}.

The pairing of perturbations can be understood by considering the symmetry of the torus\footnote{We thank Jonathan Rawlinson for showing us this observation and calculation, and for generously allowing us to reproduce our own version of it here.}. Consider attractive channel scattering, as seen in figure \ref{fig:att}. The reflection group $G$ is a symmetry group along the entire path, and in particular it is a subgroup of $D_{\infty h}$, the symmetry group of the torus. Hence a perturbation which transforms as an irrep of $G$ transforms as a restricted representation of $D_{\infty h}$. Consider the perturbation $(+-+)$. There are no one-dimensional irreps of $D_{\infty h}$ which transform like this, meaning that the perturbation must be part of a larger irrep. There are four 2-dimensional irreps of $D_{\infty h}$; under restriction to $G$ they decompose as\footnote{We use the notation of \cite{Cotton} when discussing irreps of $D_{\infty h}$.}
\begin{align}
	E_{1g} &\to (-+-)\oplus(+--)\\ 
	E_{2g} &\to (+++)\oplus(--+) \\
	E_{1u} &\to (-++)\oplus(+-+) \\
	E_{2u} &\to (++-)\oplus(---) \, .
\end{align}
This suggests that the $(+-+)$ perturbation could be part of the $E_{1u}$ irrep. If true, there would be a partner perturbation which transforms as $(-++)$: exactly how the relative size fluctuation transforms. This suggests that the two modes are part of the same representation, explaining why they are linked through scattering. The $E_{1u}$ perturbation is also seen in numerical analyses of the Skyrme torus \cite{Gudnason:2018ysx,Barnes:1997qa}, matching the physical interpretation here. It is also geometrically natural that changing the centre of the circle should be thought of as a single deformation.

\begin{figure}[h!]
	\begin{center}
		\includegraphics[width=1.0\textwidth]{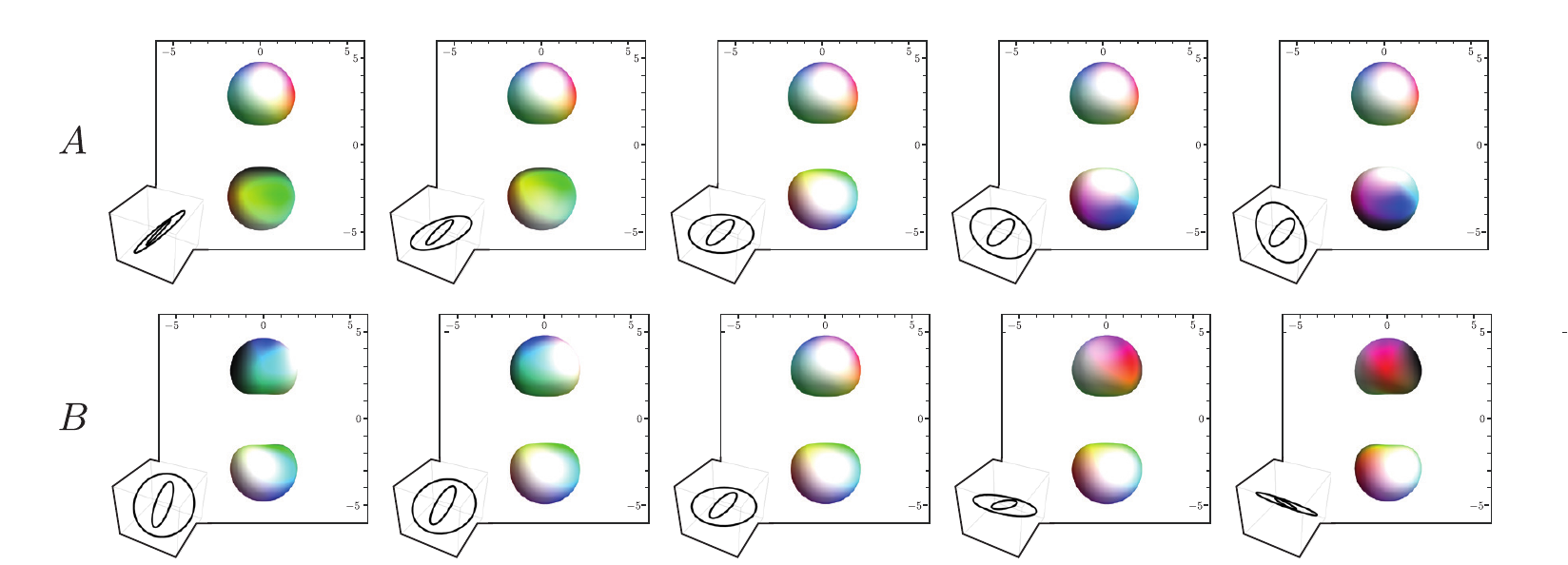} 
	\end{center}
	\caption{Rotating the 2-plane into the holographic direction. Here, we plot only the Skyrme energy density isosurface $\mathcal{E}_S$. In figure 10A, we plot configurations from $F5$ \eqref{9a} with $t_5 = \{-\pi/3, -\pi/6, 0, \pi/6, \pi/3\}$. In figure 10B, we plot configurations from $F6$ \eqref{9b} with $t_6 = \{-\pi/3, -\pi/6, 0, \pi/6, \pi/3\}$.  } 
	\label{fig:rots}%
\end{figure}

We now consider transformations which alter how the 2-plane is embedded in $\mathbb{R}^4$. The embedding in $\mathbb{R}^3_{\boldsymbol{x}}$ simply determines the overall rotational and isorotational orientation of the skyrmions. Hence why we chose our initial 2-plane to lie in the $z=0$ hyperplane, with normal vector $\boldsymbol{n} = (0,0,1) \in \mathbb{R}_x\times\mathbb{R}_y\times\mathbb{R}_h$. Rotations about this normal give rise to physical rotations around the $z$-axis. The remaining nontrivial transformation is to rotate $\boldsymbol{n}$ in the $3D$ space parametrised by $(x,y,h)$. This is a two-dimensional set of perturbations, generated by rotations about the $x$- and $y$-axes. These motions physically affect the Skyrme and meson fields in different ways; in contrast to the modes discussed above. We will attempt to separate these effects by first only plotting the Skyrme energy density $\mathcal{E}_S$. Consider a rotation about the major axis of the ellipse, which coincides with the line joining the skyrmions. We take the family
\begin{equation} \label{9a}
	F5: e_1 = 3, \quad e_2 = 1, \quad c_1 = c_2 = 0, \quad \boldsymbol{n} = \left(\sin(t_5),0,\cos(t_5)\right) \, .
\end{equation}
 We plot the corresponding Skyrme energy densities in figure \ref{fig:rots}A. The skyrmions change their relative phase and move out of the attractive channel while retaining a reflection symmetry $x \to -x$. By studying the Skyrme energy density, we find that this perturbation transforms as $(+--)$.  Now consider a rotation about the minor axis of the ellipse. This moves the major axis of the ellipse into the $yh$-plane. The position of the skyrmions are given by the projection of the vertices of the ellipse onto the $y$-axis. Hence we modify the length of the major axis to keep the skyrmion positions fixed. In detail the family is given by
\begin{equation} \label{9b}
	F6: e_1 = 3/\cos(t_5), \quad e_2 = 1, \quad c_1 = c_2 = 0 ,\quad \boldsymbol{n} = (0,\sin(t_6),\cos(t_6))\, .
\end{equation} 
We plot the corresponding Skyrme energy densities in figure \ref{fig:rots}B, and again the skyrmions change phase. The perturbation transforms as $(-+-)$. However, the skyrmions remain in the attractive channel. When only looking at the Skyrme energy density, this transformation looks like a zero mode: an isorotation about the $u_2$-isoaxis. We can already describe this mode using a global gauge transformation. Hence, it appears that the map from instantons to skyrmions is not injective. To reintroduce injectivity, we must reintroduce mesons.

\begin{figure}[h!]
	\begin{center}
		\includegraphics[width=1.0\textwidth]{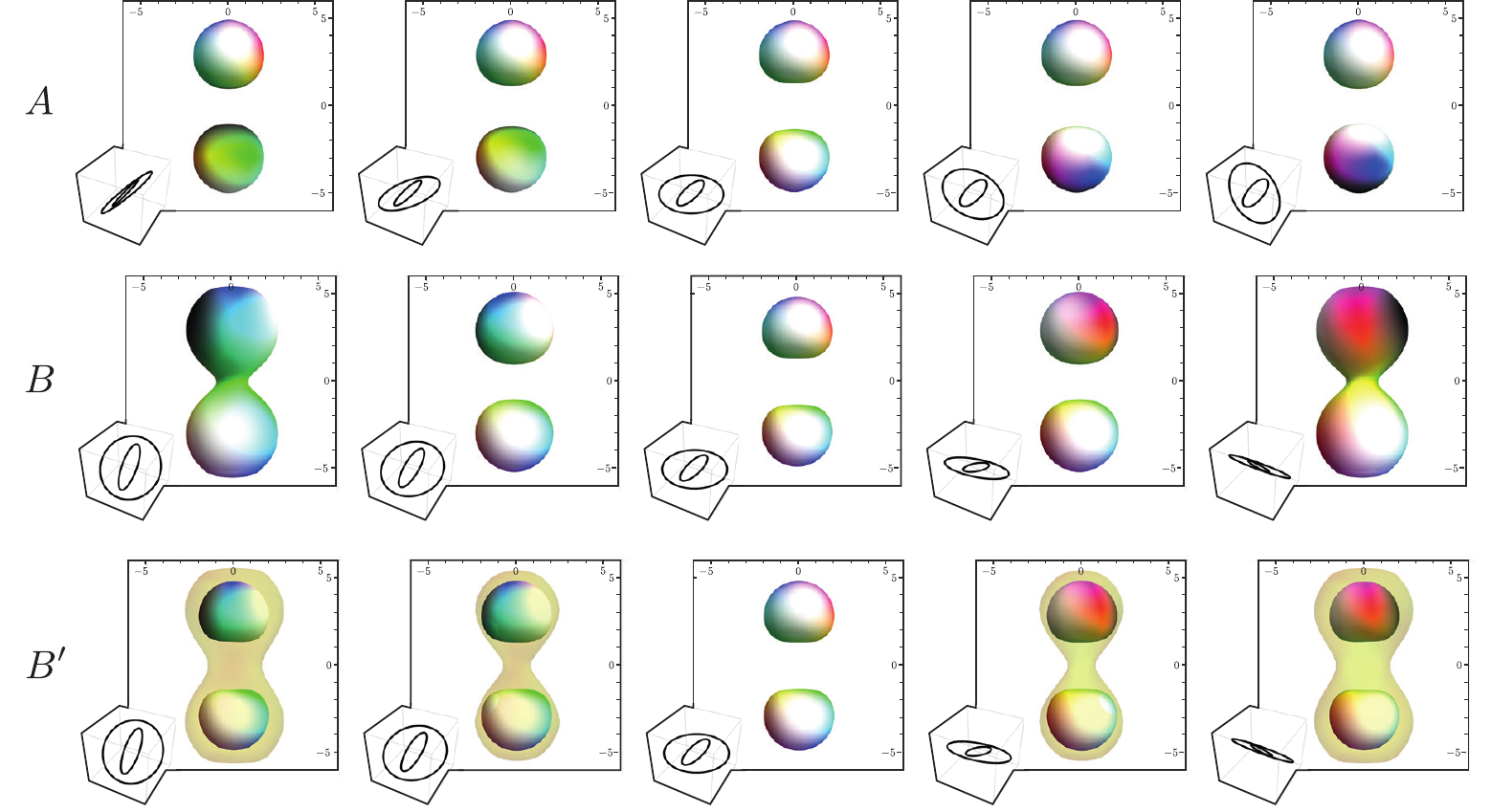} 
	\end{center}
	\caption{Rotating the 2-plane into the holographic direction. In contrast to figure \ref{fig:rots}, we now plot an isosurface of the entire energy density, $\mathcal{E}_S + \mathcal{E}_V + \mathcal{E}_I$. In the bottom row, we plot two isosurfaces: one of the Skyrme energy density $\mathcal{E}_S$ and one of remaining energy density $\mathcal{E}_V+\mathcal{E}_I$. In figure 12A, we plot configurations from $F5$ \eqref{9a} with $t_5 = \{-\pi/3, -\pi/6, 0, \pi/6, \pi/3\}$. In figure 11B and 11B', we plot configurations $F6$ \eqref{9b} with $t_6 = \{-\pi/3, -\pi/6, 0, \pi/6, \pi/3\}$. } 
	\label{fig:incmes}%
\end{figure}

Let us reconsider the families $F5$ and $F6$, now including the meson energy densities $\mathcal{E}_V + \mathcal{E}_I $. In the $F6$ family the JNR data rotates around the $x$-axis, moving the major axis of the ellipse into the $yh$-plane. Each instanton is approximately positioned on the major axes of the ellipse, so the perturbation moves the instantons away from $h=0$. Intuition from the $N=1$ sector suggests that the corresponding meson field's energy should increase. In contrast, the instantons in the family $F5$ stay near $h=0$, and hence the meson field should not significantly change. We can test this hypothesis by considering the Skyrme and meson energy densities, $\mathcal{E}_S$ and $\mathcal{E}_V + \mathcal{E}_I$, seperately. We find that $\mathcal{E}_V + \mathcal{E}_I$ is approximately invariant for mode $F5$ but changes significantly for mode $F5$. To see this, we plot the full energy densities $\mathcal{E}_S +\mathcal{E}_V + \mathcal{E}_I$ for families $F5$ and $F6$ in figures \ref{fig:incmes}A and B. We also plot an isosurface of $\mathcal{E}_V + \mathcal{E}_I$in \ref{fig:incmes}B'. Here we see that the meson energy grows as the mode is excited, confirming our hypothesis. Of course, any separation of the Skyrme and meson densities is somewhat arbitrary, since the fields are strongly coupled.
 
\begin{figure}[h!]
	\begin{center}
		\includegraphics[width=1.0\textwidth]{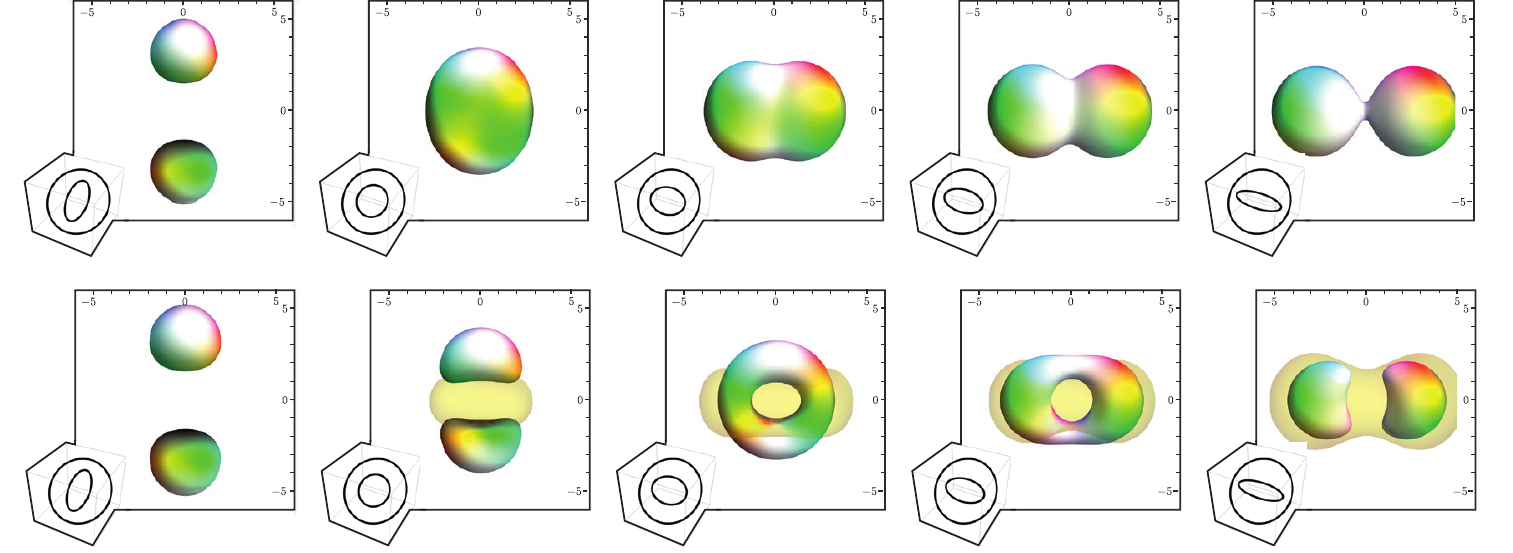} 
	\end{center}
	\caption{A scattering path which exchanges orientational energy for meson size. The configurations are given by the family \eqref{11}, with $t_7 =  \{ \pi/4,5\pi/12,\pi/2,7\pi/12,3\pi/4 \}$. We plot surfaces of constant energy density $\mathcal{E}_S + \mathcal{E}_V + \mathcal{E}_I$ in the top row. We plot the same family in the bottom row, but plot two surfaces of constant density: the Skyrme energy density $\mathcal{E}_S$ and the remaining energy density $\mathcal{E}_V+\mathcal{E}_I$. }
	\label{fig:messca}%
\end{figure}

The perturbations described by $F5$ and $F6$ are linked through scattering. To show this, we start with a configuration out of the attractive channel (by exciting $F5$) and perform a scattering. Such a path is given by
\begin{align} \label{11}
	F7: e_1 = 5/2 - 3/2\cos(t_7), \quad e_2 = 5/2 + 3/2\cos(t_7) ,\nonumber \\
	c_1 = c_2 = 0, \quad \boldsymbol{n} = \left(0.7, -\sqrt{1-0.7^2},0\right)\, .
\end{align}
The corresponding energy densities are plotted in figure \ref{fig:messca}. Note that the outgoing skyrmions are in the attractive channel. Physically, the initial configuration has a high potential energy due to skyrmion orientation. After scattering the skyrmions enter the attractive channel, so the orientational potential energy is transferred to the meson field. This observation should have significant consequences for nucleon scattering. Certain nucleon-nucleon polarisations, which excite this orientational mode, will be more likely to produce excited vector mesons. At infinite separation the orientational mode is a zero mode, while the meson-size mode is not. The $E_{1g}$ mode decomposes as $(-+-)\oplus(+--)$ under restriction to $G$. Hence the modes are part of a 2-dimensional family which transform as $E_{1g}$.

\begin{figure}[h!]
	\begin{center}
		\includegraphics[width=1.0\textwidth]{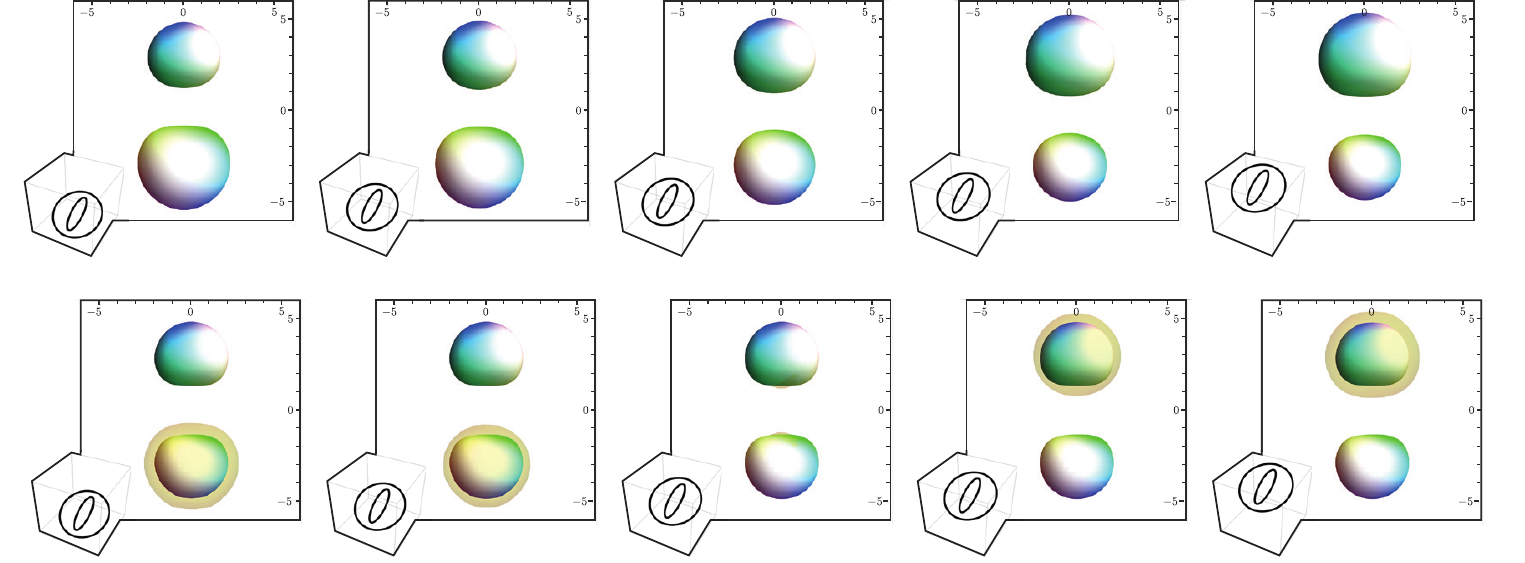} 
	\end{center}
	\caption{The fluctuation where the relative meson field size changes. The configurations are given by the family \eqref{eq:relative}, with $t_8 =  \{ -2.5,-1.0, 0, 1, 2.5 \}$.  We plot surfaces of constant energy density $\mathcal{E}_S + \mathcal{E}_V + \mathcal{E}_I$ in the top row. We plot the same family in the bottom row, but plot two surfaces of constant density: the Skyrme energy $\mathcal{E}_S$ and the remaining energy density $\mathcal{E}_V+\mathcal{E}_I$.. }
	\label{fig:relative}%
\end{figure}

Finally, we consider how to generate a configuration where the mesons attached to each skyrmion can have different energies. Suppose that the instantons are positioned on the vertices of the major axis of the ellipse. Then in the data we've studied so far, the instantons have equal positions (up to a sign) on the $h$-axis. This retains a symmetry in the meson field. To break the symmetry we excite the meson size mode (as in $F6$) while moving the 2-plane in the $h$-direction. Doing either motion individually retains symmetry between the two meson fields but the combination does not. Consider the family
\begin{equation} \label{eq:relative}
F8: e_1 = 3, e_2 = 1, \quad c_1 = c_2 = 0, \quad \boldsymbol{n} = (0.7, -\sqrt{1-0.7^2}, 0) \, ,
\end{equation}
where the 2-plane is then shifted in the $h$-direction by $t_8$. The corresponding family is plotted in figure \ref{fig:relative}, and the relative energy of the meson fields surrounding each skyrmion change with $t_8$. We also plot the Skyrme energy density $\mathcal{E}_S$ and the remaining energy density $\mathcal{E}_V + \mathcal{E}_I$. Here, it is clear that the Skyrme energy does not noticeably change size while the meson field does

We have now interpreted the $N=2$ instanton moduli as physical degrees of freedom of the two skyrmion system coupled to a meson field. We now summarise the results. The instanton data is given by an ellipse and circle embedded on a 2-plane in $\mathbb{R}^4$. The axes of the ellipse $(e_1,e_2)$ control the overall size and separation of the skyrmions. The centre of the circle $(c_1,c_2)$ control one relative orientation and the relative size of the skyrmions, as demonstrated by the families $F2$ and $F3$ displayed in figure \ref{fig:circ1}. The meson field becomes relevant when we consider the embedding of the 2-plane in $\mathbb{R}^4$. By rotating the 2-plane away from the $h=0$ hyperplane we can excite two more modes: another relative orientational mode and the meson energy mode. These modes are captured by families $F5$ and $F6$ and are displayed in figure \ref{fig:incmes}. Finally, the relative meson energy fluctuation can be generated by also translating the 2-plane in the $h$-direction.

We have presented an analysis of perturbations around a simple initial configuration of two separated skyrmions in the attractive channel. However our physical interpretation should apply more globally, though this requires a more thorough analysis. One immediate consequence is a mode analysis around the toroidal skyrmion. We predict that the non-zero modes decompose as
\begin{equation} \label{eq:spectrum}
	A_{1g} + E_{2g} + E_{1u} +  E_{1g} \, ,
\end{equation}
where the modes physically correspond to: an overall size fluctuation ($A_{1g}$); $90^\circ$ scattering ($E_{2g}$); relative orientation/size fluctuations ($E_{1u}$) and relative orientation/meson-energy fluctuations ($E_{1g}$). Note that this decomposition was not seen in the pure Skyrme theory \cite{Gudnason:2018ysx, Barnes:1997qa}. Hence we must study a model which includes mesons to have a chance of reproducing the result \eqref{eq:spectrum}. Houghton has performed a mode analysis of the $N=3$ tetrahedral instanton-generated skyrmion \cite{Houghton:1999uq}. The predicted mode decomposition is
\begin{equation} \label{eq:houghton}
	A_1 + 2E + F_1 + 2F_2 \, ,
\end{equation}
giving another non-trivial check of the model. To reemphasise: the result \eqref{eq:houghton} is not seen for pure Skyrme theory, it should only appear in a theory with mesons.

We finish this section by considering the global picture. The 16-dimensional $N=2$ instanton moduli space is isomorphic to \cite{Atiyah:1992if}
\begin{equation}
	C_8 \times \mathbb{R}^4 \times \mathbb{R}_{>0} \times SU(2) \, ,
\end{equation}
where $\mathbb{R}^4$, $\mathbb{R}_{>0}$ and $SU(2)$ represent the translations, scaling and global gauge transformations respectively. The remaining 8-dimensional space $C_8$ is the manifold of Hartshorne ellipses. The space is isomorphic to $\mathbb{C}P^4$ with real points removed and conjugate points identified. This space still contains two zero modes; the remaining rotations. There then remains a six-dimensional non-trivial space of configurations with different energies. On $C_8$, transformations which are important for quantum mechanics (such as the interchange of two particles) have simple realisations, hence it is a nice space to quantise on. Atiyah and Manton constructed a six-dimensional subspace of $C_8$ as part of a 12-dimensional skyrmion configuration space which included well-separated skyrmions. Our analysis shows that this cannot model low energy dynamics as it does not contain enough degrees of freedom to describe the full scattering paths shown in figures \ref{fig:circ2} and \ref{fig:messca}. To include these paths, we should consider the entire space $C_8$. This is mathematically simpler than constructing a subspace, but leads to a higher-dimensional model which will be challenging to study computationally.  

\section{Further Work and Conclusion}

We have given the first construction of a manifold of low energy 2-skyrmions which includes enough degrees of freedom to describe well separated solutions, and their scatterings. This means our space is a Quantum Soliton Scattering Manifold (as defined in \cite{Halcrow:2020lez}) and can be used to describe nucleon-nucleon scattering. To proceed further, one must calculate the potential and metric on our proposed manifold. This may sound daunting as the configuration space is 16-dimensional, but the metric and potential only vary in the centre of the manifold. One might be able to consistently fix an optimal size for the meson and Skyrme fields at each point, leaving only a five-dimensional non-trivial subspace. We interpret the non-trivial modes as controlling the relative separation, orientation, size and meson-energy of the configuration. If this manifold can be constructed, one could study nucleon-nucleon scattering within the Skyrme model: an open problem since Skyrme wrote down the model in 1961 \cite{Skyrme1961}.

Past attempts to construct a two-skyrmion configuration space have run into problems. Our work suggests that the inclusion of a single vector meson resolves these issues, since the instanton moduli space generates a consistent skyrmion configuration space. Without the mesons, the degrees of freedom in the instanton model do not uniquely translate to those in the Skyrme model. Recently Komargodski \cite{komargodski2019baryons} proposed a new way of understanding baryons in $N_f = 1$ QCD. Karasik then showed that mesons are the key to understanding the transition between baryons as skyrmions when $N_f=2$ and baryons as described by Komargodski when $N_f=1$ \cite{Karasik:2020pwu, karasik2020vector}. Hence there is growing evidence that the pure Skyrme model is fundamentally insufficient; but that this is remedied by the inclusion of vector mesons.

Our configuration space is in some sense minimal. To describe well separated skyrmions and their zero modes, one must include fluctuations which change the relative orientations of the skyrmions. But these fluctuations are intrinsically linked to the relative size and meson-energy fluctuations, due to the mode-pairing discussed in section 4. Hence one must include these degrees of freedom to have a complete picture of the two-skyrmion space.

A key advantage of our construction is that it generalises to any baryon number. The low-energy $N$-skyrmion configuration space, when coupled to a vector meson, is described by the $N$-instanton moduli space. Although this map from instantons to skyrmions is not well studied for $N>2$, it provides a systematic framework for the study of skyrmions. To proceed for higher $N$ it is important to understand how common ideas in instanton theory, such as ADHM data \cite{Atiyah:1978ri}, translate to the Skyrme model. This has been only studied in highly symmetric cases  \cite{Leese:1993mc, Singer:1999rx}.

We have analysed the instanton approximation of the model \eqref{eq:toteng}. While there is good evidence to trust this approximation, it would be good to test the results numerically. There are several checks to do. One could excite an orientational mode of the skyrmions when well separated and boost them towards one another. This would test to see if the scattering paths \ref{fig:circ2} and \ref{fig:messca} are a genuine feature of the Skyrme model defined by \eqref{eq:energies}, rather than an artefact of the instanton approximation. One could also calculate the normal mode spectrum around the toroidal skyrmion and compare it to \eqref{eq:spectrum}.

One key remaining problem is the pion mass. Instanton-generated skyrmions have an energy density which falls off polynomially in $r$. This matches the polynomial decay of skyrmions with massless pions. However, pions do have mass and physically realistic Skyrme fields must decay exponentially. Hence, to make a connection with nuclear physics, one must somehow include this fact. It has been demonstrated that one can generate spherical skyrmions with exponential decay from skyrmions in hyperbolic space \cite{Atiyah:2004nh}. However, it is not clear how to generalise this work to separated skyrmions.

The model \eqref{eq:energies} as derived directly from Yang-Mills theory is rather complicated, with six interaction terms. The model is chosen to have a low classical binding energy, although we note that this is not directly comparable with the quantum binding energy as measured in nuclear physics. We have demonstrated that a mathematically consistent 2-skyrmion space requires the inclusion of a meson. One may ask: what is the simplest way to couple a meson to the Skyrme model? And, can we modify the interaction to better match meson physics? To help answer these questions, one could include vector mesons in a nucleon-nucleon interaction calculation, adapting the formalism developed in \cite{HH2020}. One could then constrain the parameters by comparing the skyrmion interaction to phenomenological nucleon-nucleon interactions.

\section*{Acknowledgements}

The authors would like to thank Jonathan Rawlinson and Derek Harland for many insightful conversations about this project, Paul Sutcliffe for some numerical reassurance and Andrzej Wereszczy\'nski and Carlos Naya for finding some errors in the manuscript. CH is supported by The Leverhulme Trust as an Early Career Fellow. TW is supported by the University of Leeds as an Academic Development Fellow. This work was supported in part by the UK Engineering and Physical Sciences Research Council through grant EP/P024688/1. All simulations were run using the Soliton Solver library developed by TW at the University of Leeds.

\bibliographystyle{ieeetr}

\end{document}